\documentclass[epj]{webofc}
\usepackage[varg]{txfonts} 
\usepackage{amssymb}
\newcommand{\beq}{\begin{equation}}
\newcommand{\eeq}{\end{equation}} 
\newcommand{\beqa}{\begin{eqnarray}} 
\newcommand{\eeqa}{\end{eqnarray}}

\def\bc{\begin{center}}
\def\ec{\end{center}}
\def\bi{\begin{itemize}}
\def\ei{\end{itemize}}
\def\be{\begin{equation}}
 \def\ee{\end{equation}}
\def\ben{\begin{equation*}}
 \def\een{\end{equation*}}
 \def\bea{\begin{eqnarray}}
 \def\eea{\end{eqnarray}}
 \def\bean{\begin{eqnarray*}}
 \def\eean{\end{eqnarray*}}
\newcommand{\ie}{{\em i.e.}}  
  \newcommand{\etal}{{\em et al.}}

\def\xtwo{x_{2}}

\newcommand{\pPb}{p--Pb}

\newcommand{\eq}[1]{(\ref{#1})}
\newcommand{\ed}{\end{document}}

\newcommand{\qhat}{\hat{q}}

\newcommand{\Phat}{\hat{{\cal P}}}
\newcommand{\epsa}{\varepsilon}
\newcommand{\epsamax}{\varepsilon^{\rm max}}
\newcommand{\sqrts}{\sqrt{s}}
\newcommand{\Ea}{E}

\newcommand{\Ep}{E_\mathrm{p}}

\newcommand{\mT}{M_\perp}

\newcommand{\tf}{t_f}

\newcommand{\jpsi}{{\mathrm J}/\psi}
\newcommand{\xf}{x_{\mathrm{F}}}
\newcommand{\xh}{x_{\mathrm{h}}}

\newcommand{\pt}{p_{_\perp}}

\newcommand{\pp}{p--p}
\newcommand{\pA}{p--A}

\newcommand{\hi}{A--A}
\newcommand{\dd}{{\rm d}}
  
\newcommand{\lsim}{\lesssim} \newcommand{\gsim}{\gtrsim}

\newcommand{\qzero}{\hat{q}_0}

\newcommand{\gevsqfm}{GeV$^2$/fm}

 \def\sig{\sigma}

 \def\esim{\,\mathrel{\rlap{\lower0.2em\hbox{$-$}}\raise0.15em\hbox{\footnotesize $\hskip0.04em\sim$}}\,}
 \def\gsim{\mathrel{\rlap{\lower0.2em\hbox{$\sim$}}\raise0.2em\hbox{$>$}}}
 \def\ksim{\mathrel{\rlap{\lower0.2em\hbox{$\sim$}}\raise0.2em\hbox{$<$}}}

\woctitle{Physics Opportunities at an Electron-Ion Collider}

\begin{document}
\title{Nuclear suppression in \pA\ collisions from induced radiation}

\author{Fran\c{c}ois Arleo\inst{1}
\and Rodion Kolevatov\inst{2}
\and St\'ephane Peign\'e\inst{3,}\footnote{Talk given at the 6th International Conference on Physics Opportunities at an Electron-Ion Collider (POETIC 2015).} 
\and Taklit Sami\inst{3}
}

\institute{Laboratoire Leprince-Ringuet (LLR), \'Ecole polytechnique, CNRS/IN2P3 91128 Palaiseau, France 
\and 
Department of Physics, University of Oslo, PB1048 Blindern, 0316 Oslo, Norway
\and SUBATECH, UMR 6457, Universit\'e de Nantes, Ecole des
Mines de Nantes, CNRS/IN2P3 \\ 4 rue Alfred Kastler, 44307 Nantes cedex 3, France }

\abstract{The current status of coherent energy loss is reviewed, both in theory and in its phenomenological applications to \pA\ collisions.}
\maketitle
\vspace{-5mm}
\section{Introduction}
\label{intro}

The nuclear suppression of hadron production in proton-nucleus (\pA) with respect to proton-proton (\pp) collisions  has been widely discussed, but is still an open question due to the presence of various competing effects, depending on the precise kinematics and collision energy. Until quite recently, most of the phenomenological approaches assumed hadron nuclear suppression to be due either to hadron nuclear absorption or to the shadowing of the gluon distribution in the target nucleus expected at small $\xtwo \lesssim 10^{-2}$. However another fundamental phenomenon, namely coherent parton energy loss in cold nuclear matter, could play a decisive role in hadron nuclear suppression~\cite{Arleo:2010rb}. 

Medium-induced coherent gluon radiation is a QCD prediction which has now been obtained in various theoretical setups~\cite{Arleo:2010rb,Arleo:2012rs,Armesto:2012qa,Armesto:2013fca,Peigne:2014uha,Liou:2014rha,Peigne:2014rka}, and was successfully applied to the phenomenology of quarkonium nuclear suppression in \pA\ collisions~\cite{Arleo:2012hn,Arleo:2012rs,Arleo:2013zua}. Here we review the current status of coherent energy loss in theory (Sect.~\ref{sec:theo}) and in its phenomenological applications to \pA\ collisions (Sect.~\ref{sec:pheno}). 

\section{Theory of induced energy loss: a brief history}
\label{sec:theo}

It was first noted by Gavin and Milana~\cite{Gavin:1991qk} that the strong increase of $\jpsi$ suppression with $\xf$ observed in \pA\ collisions could be easily reproduced by {\it assuming} an induced parton energy loss $\Delta E$ scaling as the energy $E$ of the energetic $c {\bar c}$ pair, namely $\Delta E \propto E$.
However, the Gavin-Milana `explanation' was soon put aside, because of a claim that any {\it induced} energy loss $\Delta E$ should be {\it bounded} in the $E \to \infty$ limit. It is now understood (see for instance Ref.~\cite{Peigne:2008wu}) that the latter claim is incorrect, more precisely the induced energy loss is not bounded in general, but only in the specific situation where the energetic parton is suddenly accelerated (as in deep inelastic scattering) in the nuclear medium. In the situation where the parton is `asymptotic', \ie\ `prepared' at $t=-\infty$ and `tagged' at $t=+\infty$ after crossing a nuclear medium of thickness $L$ (a situation relevant to forward hadron production in \pA\ collisions), the behavior $\Delta E \propto E$ is correct~\cite{Arleo:2010rb}. Below we briefly review the features of induced radiation in these two different situations. 

\subsection{Parton suddenly produced in a medium}
\label{sec:case1}

\begin{figure}[t]
\centering
\includegraphics[width=9cm]{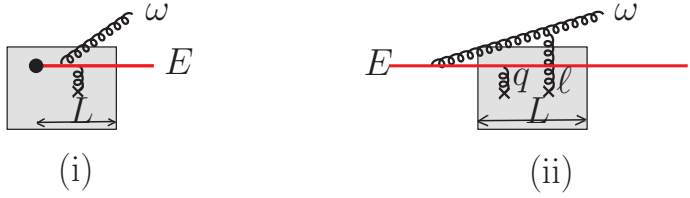}
\caption{Two qualitatively different situations relevant to medium-induced parton energy loss: (i) parton suddenly accelerated in a medium and (ii) `asymptotic parton' crossing a medium.}
\label{fig-1}      
\end{figure}

\begin{figure}[t]
\centering
\includegraphics[width=7cm]{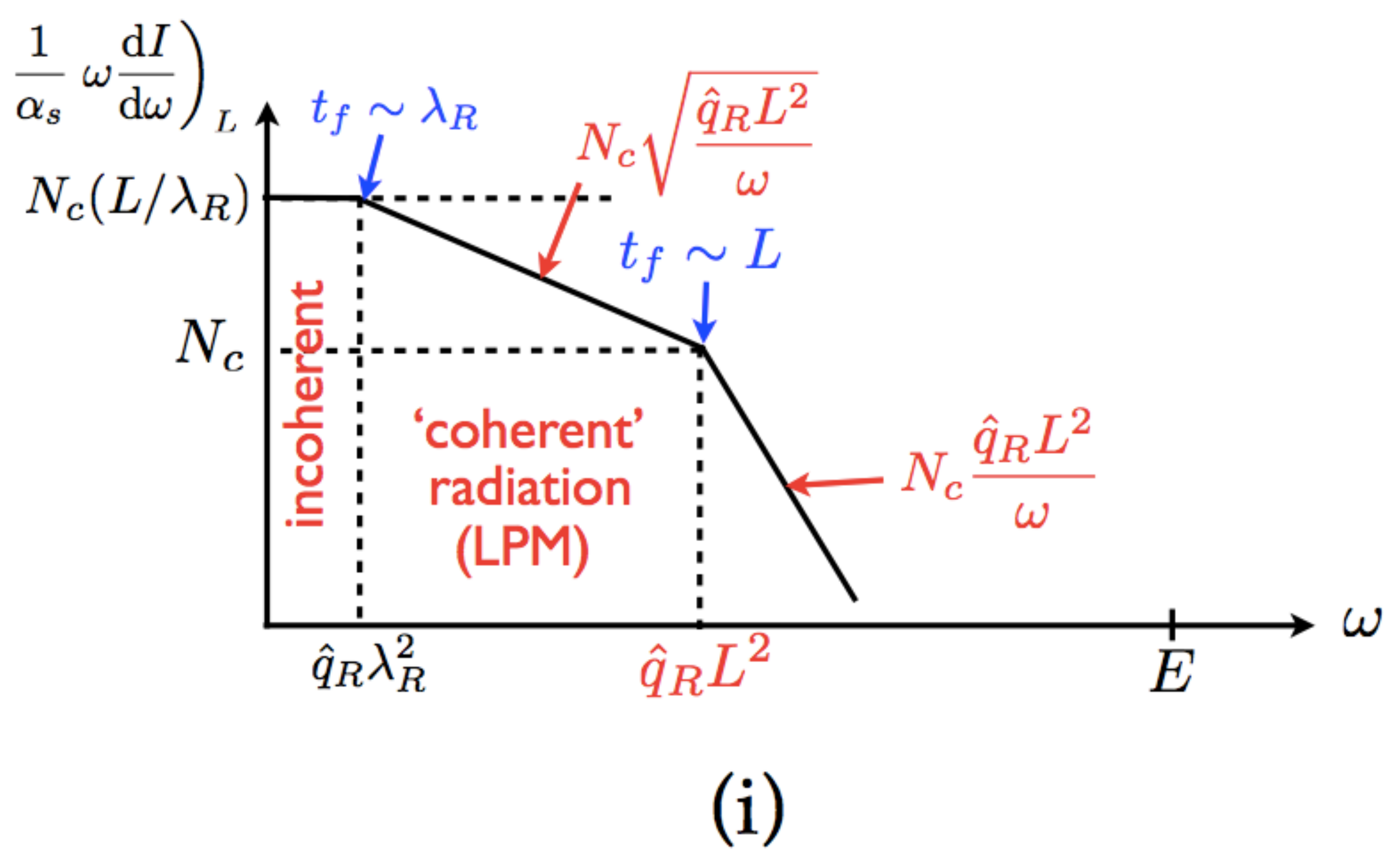} \includegraphics[width=7cm]{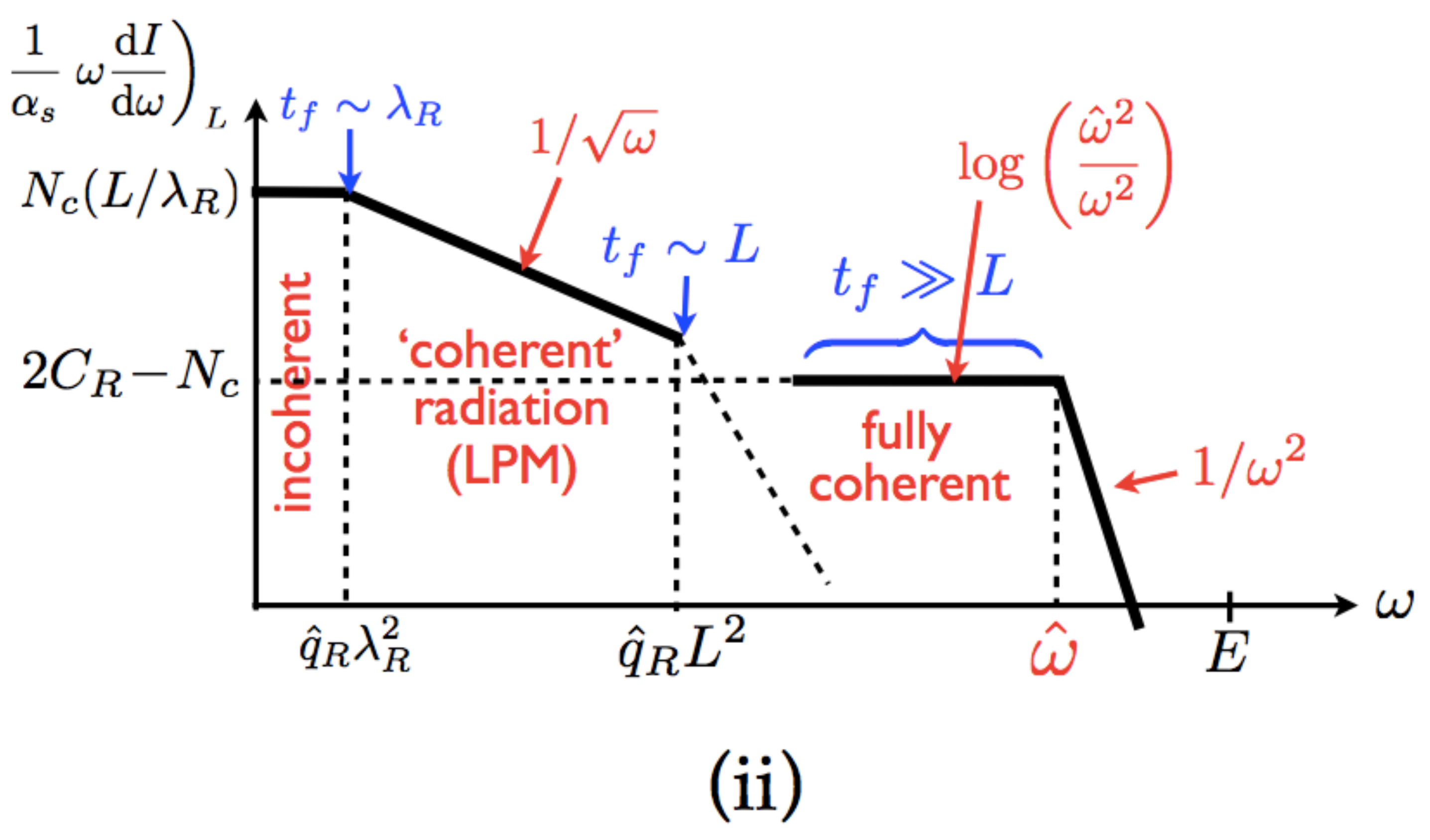}
\caption{Log-log schematic plot showing the parametric dependence of the induced radiation spectrum $\omega \dd I / \dd \omega$ (divided by $\alpha_s$), in case (i) of a parton produced in a medium (left) and in case (ii) of an `asymptotic parton' (right). See text for details.}
\label{fig-2}      
\end{figure}

The case (i) of a parton suddenly created in a medium is sketched in Fig.~\ref{fig-1} (left).
The associated medium-induced (soft) gluon radiation spectrum $\omega \dd I / \dd \omega$ has been calculated before and is represented schematically in Fig.~\ref{fig-2} (left). 

It is useful to recall the different regimes identified in~\cite{Baier:1996kr} depending on the value of the gluon formation time $\tf = \omega/k_\perp^2$ (for a more detailed discussion in QED and in QCD, see~\cite{Peigne:2008wu}). In the {\it incoherent} or {\it Bethe--Heitler} regime $\tf < \lambda_R$ (where $\lambda_R$ is the parton mean free path in the medium), corresponding to very soft gluons $\omega < \hat{q}_R \lambda_R^2$ (where $\hat{q}_R \equiv \mu^2/\lambda_R$ with $\mu$ the typical transverse momentum exchange in a single scattering), each scattering center acts as an independent source of radiation. In this regime the energy spectrum $\omega \dd I / \dd \omega$ is independent of $\omega$ (up to logarithms). In the domain $\lambda_R < \tf < L$, corresponding to intermediate radiated energies $\hat{q}_R \lambda_R^2 < \omega < \hat{q}_R L^2$, a group of $\sim \tf / \lambda_R$ scattering centers acts as a single radiator, leading to a relative suppression $\propto 1/\sqrt{\omega}$ of the gluon radiation spectrum with respect to the Bethe-Heitler regime~\cite{Baier:1996kr,Zakharov:1997uu}. This is the so-called Landau--Pomeranchuk--Migdal (LPM) suppression regime. Finally, in the domain of large formation time $\tf > L$ (or $\omega > \hat{q}_R L^2$), called the {\it fully coherent} domain in the following, all scattering centers in the medium act coherently as a source of radiation, and the induced radiation spectrum behaves as $\sim 1/\omega$~\cite{Zakharov:2000iz}, \ie\ is more strongly suppressed than in the LPM regime. 

In the situation (i), the suppression of the radiation spectrum in the fully coherent domain can be simply understood as follows. For formation times $\tf \gg L$, the radiation spectrum must arise dominantly from time-ordered diagrams where the gluon emission vertex is far beyond the medium. The associated gluon emission amplitude is then $\propto (\vec{\theta}-\vec{\theta}_s)/(\vec{\theta}-\vec{\theta}_s)^2$~\cite{Arleo:2010rb}, where $\vec{\theta} \equiv \vec{k}_\perp/\omega$ and $\vec{\theta}_s \equiv \vec{p}_\perp/E$ are the final `angles' of the radiated gluon and energetic parton, respectively. The double differential spectrum $\omega \dd I / \dd \omega \dd^2 \vec{\theta} \propto 1/(\vec{\theta}-\vec{\theta}_s)^2$ depends on the medium only through $\vec{\theta}_s$, which is related to the medium size by the random walk estimate ${\theta}_s^2 \simeq (\mu^2/E^2) \cdot (L/\lambda_R)$ resulting from multiple scattering of the energetic parton across the medium. The spectrum {\it integrated over gluon angles}, $\omega \dd I / \dd \omega \sim \int \dd^2 \vec{\theta}/ (\vec{\theta}-\vec{\theta}_s)^2$, is independent of $\vec{\theta}_s$ (by a trivial change of variable in the $\vec{\theta}$-integral) and thus of $L$, leading to a strong suppression of the {\it induced} energy spectrum. The above heuristic discussion illustrates why large formation times $\tf \gg L$ are suppressed in situation (i). 

Due to this suppression, the spectrum decreases as $1/\omega$ when $\omega > \hat{q}_R L^2$~\cite{Zakharov:2000iz}, leading to an {\it average} energy loss in situation (i) effectively dominated by $\omega \sim \hat{q}_R L^2$ and given by (up to logarithms)
\be
\label{LPM-loss}
\Delta E \equiv \int \dd \omega \, \omega \frac{\dd I}{\dd \omega} \sim \alpha_s N_c \, \hat{q}_R L^2 = \alpha_s C_R \,\hat{q} L^2 \, ,
\ee
where $C_R$ is the energetic parton color charge, and $\hat{q} \equiv \hat{q}_g = \mu^2/\lambda_g$ the {\it gluon} transport coefficient. For a rigorous derivation of the induced radiation spectrum and associated average energy loss in situation (i), see Refs.~\cite{Baier:1996kr,Zakharov:1997uu}. 

\subsection{Fully coherent energy loss}
\label{sec:case2}

We now turn to the case (ii) of an asymptotic parton (see Fig.~\ref{fig-1}, right), which is relevant to forward hadron production in \pA\ collisions. The fundamental difference with the case (i) reviewed in the previous section is that large formation times $\tf \gg L$ are not suppressed any longer, but on the contrary {\it dominant} in $\omega \dd I / \dd \omega$. This can be simply understood as follows. Supposing $\tf \gg L$, three types of contributions to the radiation spectrum can emerge: purely final state radiation, where the gluon emission vertex is associated with a large {\it positive} time $t \gg L$ in both the amplitude and its conjugate, purely initial state radiation corresponding to large {\it negative} emission times in the amplitude and its conjugate, and an interference contribution where the gluon is emitted with $t \gg L$ in the amplitude and $-t \gg L$ in the conjugate amplitude (or vice versa). The two former contributions are suppressed in the {\it induced} spectrum, for the same reason as in the case (i), see Sect.~\ref{sec:case1}. On the contrary, the latter interference  \vspace{-1mm} contribution to $\omega \dd I / \dd \omega$ is of the form $\sim \int \! \dd^2\vec{\theta} \ (\vec{\theta}/\theta^2) \cdot  [(\vec{\theta}-\vec{\theta}_s)/(\vec{\theta}-\vec{\theta}_s)^2]$, and its dependence on $\vec{\theta}_s$ (and thus on $L$) cannot be removed by a simple shift in the $\vec{\theta}$-integral. It thus survives in the induced spectrum, and actually gives the dominant contribution in case (ii)~\cite{Arleo:2010rb}. Moreover, the dependence of the interference on the parton scattering `angle' $\vec{\theta}_s$ leads to an explicit dependence of the induced spectrum on the hard exchange $\vec{q}_\perp = E \vec{\theta}_s$ suffered by the parton. 

As can be seen in Fig.~\ref{fig-2} (right), the resulting induced radiation spectrum in the region $\tf \gg L$ is logarithmic, up to the scale $\hat{\omega} \equiv (\!\sqrt{\hat{q} L}/M_\perp)\,E$ (where $M_\perp \equiv (M^2 + q_\perp^2)^{1/2}$ with $M$ the parton mass), above which it is suppressed as $1/\omega^2$. The resulting average energy loss is thus dominated by $\omega \sim \hat{\omega}$ and reads~\cite{Arleo:2010rb}
\be
\label{coh-loss}
\Delta E \sim \alpha_s \, \hat{\omega} = \alpha_s \frac{\sqrt{\hat{q} L}}{M_\perp} \,E \, .
\ee
In summary, the interference contribution, which is absent in the situation (i), leads in the situation (ii) to an average parton energy loss proportional to $E$ and arising from the {\it fully coherent} domain of formation times $\tf \gg L$. 

Let us remark that for the induced radiation associated with $\tf \lsim L$, the precise way the parton is produced (in the medium or long before) does not matter. Thus, the induced spectra in the corresponding $\omega$-domain, $\omega < \hat{q}_R L^2$, are parametrically the same in both cases (i) and (ii), see Fig.~\ref{fig-2}. Note that in case (ii), this domain gives a contribution to $\Delta E$ of the LPM type \eq{LPM-loss}, but at large $E$ this contribution is negligible compared to the coherent energy loss \eq{coh-loss}. 

The rigorous calculation of medium-induced coherent radiation has been addressed by several groups, using different formalisms and considering different particular cases. Ref.~\cite{Arleo:2010rb} (see also \cite{Arleo:2012rs}) studied the induced coherent radiation associated with the hard forward $g \to Q\bar{Q}$ process (mediated by a single hard gluon exchange in the $t$-channel), the final $Q\bar{Q}$ pair being a massive {\it pointlike} color octet, in a Feynman diagram calculation and at first order in the opacity expansion formalism~\cite{Gyulassy:2000er}. Induced coherent radiation was also studied using a semi-classical method in Refs.~\cite{Armesto:2012qa,Armesto:2013fca}, at first order~\cite{Armesto:2012qa} and all orders~\cite{Armesto:2013fca} in opacity, and in a similar kinematical setup as that of Ref.~\cite{Arleo:2010rb}, however for the $q \to q$ case of a massless quark experiencing a hard scattering mediated by a {\it color singlet} exchange in the $t$-channel.
In Ref.~\cite{Peigne:2014uha}, induced coherent radiation was revisited and derived to all orders in the opacity expansion, for any $1\to 1$ hard forward process. A general expression encompassing the particular cases studied before was found for the induced (soft) coherent spectrum~\cite{Peigne:2014uha},
\be
\label{spec-general}
\omega \frac{\dd I}{\dd \omega} =  (C_R + C_{R'} - C_{t}) \, \frac{\alpha_s}{\pi} \, \log{\left(1+ \frac{\hat{\omega}^2}{\omega^2} \right) } \ ; \ \ \ \ \ \ \hat{\omega} \equiv \frac{\!\sqrt{\hat{q} L}}{\,M_\perp} \,E \, ,
\ee
with $C_R$, $C_{R'}$ the incoming and outgoing color charges, and $C_t$ the color charge of the hard $t$-channel exchange.\footnote{The color factor in \eq{spec-general} has a very simple explanation~\cite{Peigne:2014uha}. The interference term is of the form $\sim 2 T_R^a T_{R'}^a$ (where $T_R^a$ and $T_{R'}^a$ are the color generators of the incoming and outgoing parton color representations $R$ and $R'$), which can be written as $2 T_R^a T_{R'}^a = (T_R^a)^2 + (T_{R'}^a)^2  - (T_{R}^a - T_{R'}^a)^2 = C_R + C_{R'} - C_{t}$.} 
Note that the transport coefficient $\qhat$ to be used in the expression of $\hat{\omega}$ is the {\it gluon} transport coefficient $\qhat \equiv \qhat_g$, independently of the incoming and outgoing parton color states~\cite{Peigne:2014uha}.
The limiting behaviors of \eq{spec-general} in the fully coherent domain, namely, the (logarithmic) `plateau' at $\omega \ll \hat{\omega}$, and the $1/\omega^2$ suppression at $\omega \gg \hat{\omega}$, are displayed in Fig.~\ref{fig-2} (right), where for the sake of illustration the incoming and outgoing partons are chosen of the same type, $C_R=C_{R'}$. With a  color octet $t$-channel exchange, $C_t=N_c$, the height of the plateau is thus $\propto (2C_R-N_c)$, see Fig.~\ref{fig-2} (right).\footnote{In the $q \to q$ case, $C_R=C_F=(N_c^2-1)/(2N_c)$, we have $2C_F-N_c<0$ and the {\it induced} coherent energy loss is negative, see Ref.~\cite{Peigne:2014uha} for a discussion and simple interpretation of this result.}

The induced coherent radiation associated with $1 \to 2$ forward processes was also addressed (in the leading logarithm and large $N_c$ limits), for $g \to q \bar q$ and $q \to q g$ in the dipole formalism~\cite{Liou:2014rha} and for $q \to q g$ and $g \to gg$ using the opacity expansion~\cite{Peigne:2014rka}. In the leading logarithm approximation, the soft induced coherent radiation does not probe the size of the final two-parton system, and thus only depends on its {\it total} color charge. Hence the conjecture, proposed and explicitly checked for $q \to q g$ and $g \to gg$ in Ref.~\cite{Peigne:2014rka}, that the spectrum associated with $1 \to n$ hard forward processes is given by an incoherent sum of spectra associated with $1 \to 1$ processes, weighted by the probabilities $P_{R'}$ for the $n$-parton state to be produced in color representation $R'$ in the hard process. This conjecture is expected to hold for any finite $N_c$ (but only in the leading logarithm approximation)~\cite{Peigne:2014rka}. 

\section{Coherent energy loss in \pA\ collisions}
\label{sec:pheno}

Induced coherent radiation arises in the production of a single forward particle~\cite{Arleo:2010rb,Arleo:2012rs,Armesto:2012qa,Armesto:2013fca,Peigne:2014uha} and in forward dijet production~\cite{Liou:2014rha,Peigne:2014rka}, suggesting the broad relevance of coherent energy loss in forward processes.

\subsection{Quarkonium suppression from fixed-target to collider energies}
\label{sec:pheno-psi}

Coherent energy loss is expected in quarkonium production in \pA\ collisions, where typically (at sufficiently large $\sqrt{s}$) a high-energy gluon from the projectile proton is scattered to a compact color octet heavy $Q \bar{Q}$ pair~\cite{Arleo:2010rb}. 

In Refs.~\cite{Arleo:2012hn,Arleo:2012rs}, coherent energy loss is implemented as follows.  The single differential p--A production cross section as a function of the quarkonium (labelled $\psi$) energy reads
\be
\label{pheno-model}
\frac{1}{A}\frac{\dd\sigma_{\mathrm{pA}}^{\psi}}{\dd \Ea} \left( \Ea \right)  = \int_0^{\epsamax} \! \! \! \dd \epsa \, \, {\cal P}(\epsa, \Ea, \qhat L) \, \frac{\dd\sigma_{\mathrm{pp}}^{\psi}}{\dd \Ea} \left( \Ea+\epsa \right) \, ,
\ee
where $\Ea$ (respectively, $\epsa$) is the energy (respectively, energy loss) of the compact $Q \bar{Q}$ pair in the rest frame of the nucleus A. The upper limit on the energy loss is $\epsamax=\min\left(\Ea,\Ep-\Ea\right)$, where $\Ep$ is the beam energy in that frame.
The energy loss probability distribution, or \emph{quenching weight}, ${\cal P}$, is simply related~\cite{Arleo:2012rs} to the induced coherent radiation spectrum for the $g \to Q\bar{Q}$ process, which spectrum is given by Eq.~\eq{spec-general} in the particular case $C_R=C_{R'}=C_t=N_c$. The quenching weight depends on the {\it gluon} nuclear broadening $\sqrt{\qhat L}$, with $\qhat$ parametrized as~\cite{Arleo:2012rs}
\be
\label{qhat-x}
\hat{q} \equiv \hat{q}_0 \left[ \frac{10^{-2}}{\min(x_0, x_2)} \right]^{0.3} \! \! \!;   \ \ \ 
 x_0 \equiv \frac{1}{2 m_\mathrm{p} L} \, ,
\ee
where $\qzero$ is the only free parameter of the model, and $x_2 = \mT \, e^{-y}/\sqrts$ (with $y$ the quarkonium rapidity in the center-of-mass frame of an elementary proton--nucleon collision, and $M=3\,{\rm GeV}$ ($M=9\,{\rm GeV}$) for the $c\bar{c}$ ($b\bar{b}$) mass in the expression of $M_\perp$). The parameter $\qhat_0$ is determined by fitting the $\jpsi$ suppression measured by E866~\cite{Leitch:1999ea} in p--W over p--Be collisions ($\sqrt{s}=38.7$~GeV), see~\cite{Arleo:2012rs}. The obtained value is $\qzero=0.075^{+0.015}_{-0.005}$~\gevsqfm.
The p--p production cross section appearing in Eq.~\eq{pheno-model} is obtained from a fit to p--p measurements.

\begin{figure}[htb]
\centering
\vspace{-2mm}
\includegraphics[height=4.5cm]{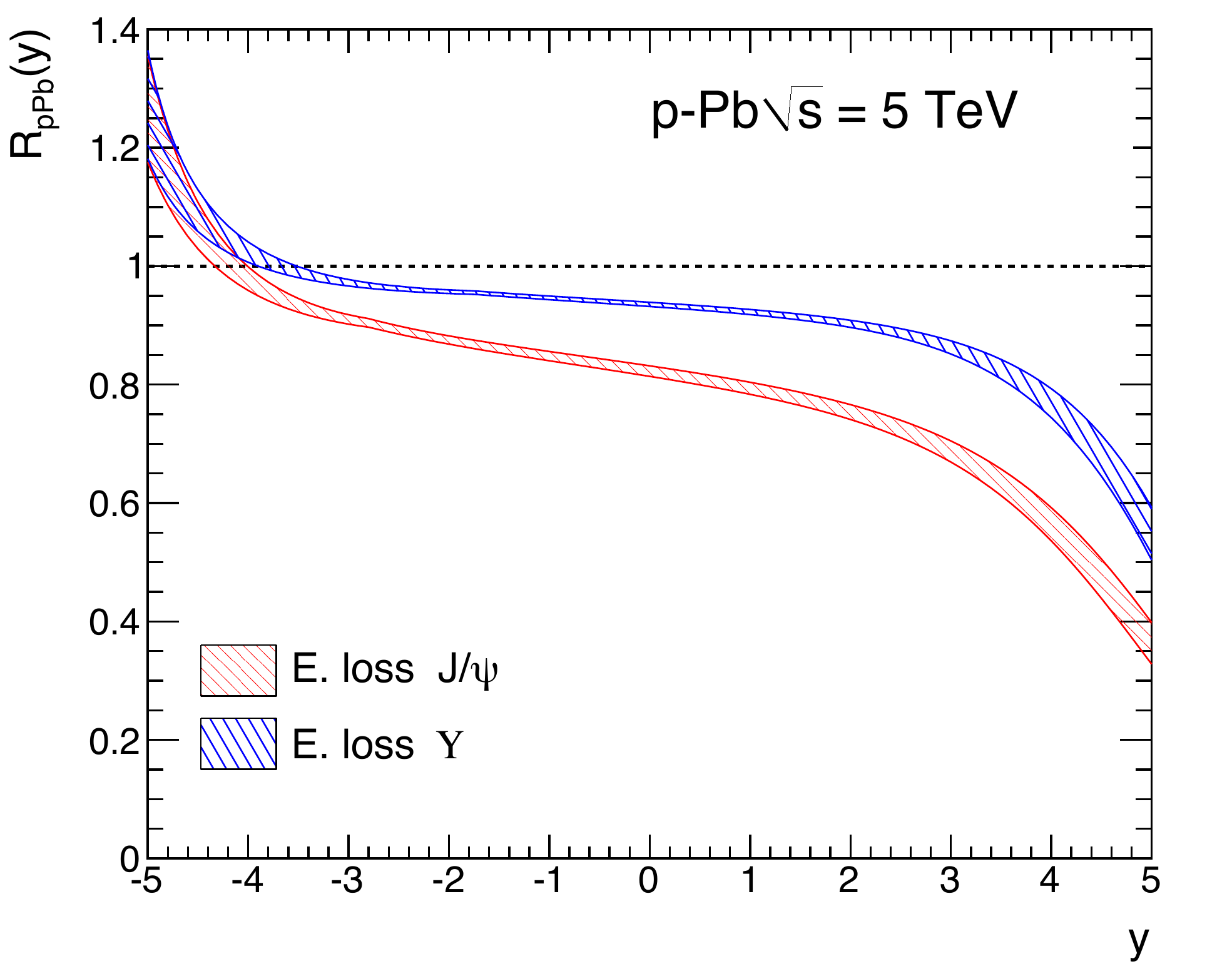}
\caption{$\jpsi$ and $\Upsilon$ suppression in p--Pb collisions at $\sqrts=5.02$~TeV from induced coherent energy loss~\cite{Arleo:2012rs}.}
\label{fig:lhc}      
\end{figure}

Quite remarkably, all available $\jpsi$ and $\Upsilon$ suppression measurements from fixed-target experiments (SPS, HERA, FNAL) to RHIC could be described within the above simple model on a broad kinematical range in rapidity $y$ (or $\xf$)~\cite{Arleo:2012rs}.\footnote{A simple extension of the model \eq{pheno-model} to double differential cross sections $\dd\sigma^{\psi}_{\rm pA}/{\dd \Ea}{\dd^2 \vec{p}_\perp}$ similarly yields a good description of the quarkonium suppression data as a function of transverse momentum~\cite{Arleo:2013zua}.} The predictions for $\jpsi$ and $\Upsilon$ suppression in p--Pb collisions at $\sqrt{s}=5.02$~TeV~\cite{Arleo:2012rs} are shown in Fig.~\ref{fig:lhc}. As can be seen, the model predicts a rather strong $\jpsi$ suppression at forward rapidity (say, $y\gtrsim3$) and a slight enhancement in the most backward rapidity bins ($y<-4$). The suppression predicted in the $\Upsilon$ channel shares the same features, however the suppression is less pronounced than that of $\jpsi$, since the (average) coherent energy loss scales as $\mT^{-1}$, see Eq.~\eq{coh-loss}. The latter predictions proved to be in excellent agreement with ALICE~\cite{Abelev:2013yxa} and LHCb~\cite{Aaij:2013zxa} data. 

\subsection{Light-hadron suppression at the LHC}

Coherent energy loss can also be applied to light-hadron production in \pA\ collisions. The picture of light-hadron production we use is as follows~\cite{AKP}. In the target rest frame, a projectile parton $i$ of high energy $E_i$ scatters on the target and produces a dijet made of two partons, $j$ and $k$, with approximately back-to-back 
transverse momenta, $|\vec{K}_{j \perp}| \simeq |\vec{K}_{k \perp}| \equiv K_{\perp}$, and energy fractions $E_j/E_i = \xh$ and $E_k/E_i = 1-\xh$. The label $j$ is chosen for the parton which then fragments (on a long timescale) into the detected hadron carrying transverse momentum $p_\perp = z\,K_\perp$ and energy $E \simeq z \, \xh \, E_i$, where $z$ is the fragmentation variable. For simplicity we shall consider only the partonic process $g\to{gg}$ (with single gluon exchange in the $t$-channel), which dominates at the LHC at not too large $\pt$. 

The dijet may be produced in different color states $R$. Since in \pA\ collisions the amount of induced coherent radiation depends on $R$ (via the global charge $C_R$), it is convenient to first express the \pp\ production cross section of a hadron with energy $E$ as an incoherent sum over color states,
\be
\label{dsig-pp-2}
\frac{\dd \sig_{\rm pp}^{h}(E)}{\dd E} = \sum_R \, \int \! \dd{\xh} \, P_R(\xh) \, \frac{\dd \sig_{\rm pp}^{h}(E, \xh)}{\dd E\,\dd{\xh}} 
\, ,
\ee
where we introduced the probability $P_R$ for the dijet to be in color state $R$ (which in general depends on the energy fraction $\xh$) and used $\sum_R P_R(\xh) = 1$. In Eq.~\eq{dsig-pp-2}, the quantity $\dd \sig_{\rm pp}^{h}(E, \xh)/\dd E\,\dd{\xh}$ is formally the cross section to find a hadron with energy $E$ in a gluon dijet with energy fractions $\xh$ and $1-\xh$. For $N_c=3$, the probability $P_R(\xh)$ for the $g\to gg$ channel vanishes in the decuplet, $P_{\bf 10 \oplus \bar{10}}(\xh) =  0$, and the $gg$ dijet can thus be color singlet, octet or vigintiseptuplet ($R={\bf 1}, {\bf 8} \equiv {\bf 8_a} \oplus {\bf 8_s}, {\bf 27}$)~\cite{AKP}. 
  
In \pA\ collisions, the medium-induced coherent radiation {\it does not probe} the dijet~\cite{Peigne:2014rka}, and thus leaves unchanged the dijet internal structure, implying that the light-cone energies of the dijet and its constituents are simply scaled down by a common factor $1 + \hat{\varepsilon}$ (with transverse momenta of the constituents and the dijet invariant mass being unchanged), where $\hat{\varepsilon} \equiv \varepsilon/E_i$ is the dijet fractional (light-cone) energy loss. Thus, for each dijet color state $R$, the \pA\ and \pp\ hadron production cross sections are related similarly to Eq.~\eq{pheno-model} in the quarkonium case. Using \eq{dsig-pp-2} we get
\be
\label{sig-pA} 
\frac{1}{A} \, \frac{\dd \sig_{\rm pA}^{h}(E)}{\dd E}  =  \sum_R  \int \! \dd \hat{\varepsilon} \, \, \hat{\cal P}_R (\hat{\varepsilon})  \! \int \! \dd \xh \, P_R(\xh) \, \frac{\dd \sig_{\rm pp}^{h}(E (1+\hat{\varepsilon}) ,\xh)}{\dd E\,\dd{\xh}}  \, , 
\ee
where the quenching weight $\hat{\cal P}_R(\hat{\varepsilon})$ is related to the induced coherent radiation spectrum corresponding to the $g \to gg$ process with the final {\it compact} gluon pair in color state $R$. The latter spectrum is given by the expression \eq{spec-general}, with $C_R \to N_c$, $C_t \to N_c$, $C_{R'} \to C_R$ (where $R={\bf 1}, {\bf 8}$, or  ${\bf 27}$), $M_\perp \to K_\perp$~\cite{Peigne:2014rka}, and $\qhat$ still given by \eq{qhat-x} but with now $x_2=p_\perp e^{-y_g}/(\xh \! \sqrt{s})$ the momentum fraction of the target gluon in the $g g \to g g$ subprocess (viewed in the proton--nucleon center-of-mass frame, $y_g$ being the rapidity in this frame of the gluon fragmenting into the hadron).  

Dividing \eq{sig-pA} by \eq{dsig-pp-2} gives the nuclear attenuation factor $R_{\rm pA}^h$. Since $P_R(\xh)$ is a rather flat function of $\xh$~\cite{Peigne:2014rka,AKP}, we approximate it as a constant in the integrand of \eq{sig-pA}, and will eventually vary $\xh$ within a realistic range in order to evaluate the associated theoretical uncertainty. Thus, 
\be
\label{RpA-E}
R^{h}_{\rm pA}(E)  \equiv \frac{1}{A} \, \frac{\dd\sigma_{\mathrm{pA}}^{h}(E)/\dd E}{\dd\sigma_{\mathrm{pp}}^{h}(E)/\dd E}  \simeq \sum_R \ P_R(\xh) \  R^{R}(E) \ ; \ \ \ R^{R}(E) \equiv  \int \! \dd {\hat{\varepsilon}} \, \, \hat{\cal P}_R (\hat{\varepsilon}) \,  \frac{\dd\sigma_{\mathrm{pp}}^{h}(E (1+\hat{\varepsilon}))/\dd E}{\dd\sigma_{\mathrm{pp}}^{h}(E)/\dd E}  \, , 
\ee
where $R^{R}(E)$ is the nuclear suppression factor of a hadron produced from a parent dijet in color state $R$. As anticipated in \cite{Peigne:2014rka}, the hadron suppression factor $R^{h}(E)$ is the average of the $R^{R}$'s over the accessible dijet color representations, weighted by the corresponding probabilities $P_R(\xh)$. 

Instead of the variables $E$ and $\hat{\varepsilon}$, we can equivalently choose the hadron rapidity $y$ (in the center-of-mass frame of the proton--nucleon collision) and the rapidity shift $\delta \equiv \ln{(1+\hat{\varepsilon})}$. 
Making the dependence on the hadron transverse momentum $p_\perp$ explicit, we can rewrite \eq{RpA-E} as 
\be
\label{RpA-y-xh}
R^{h}_{\rm pA}(y, p_\perp) \simeq \sum_R \ P_R(\xh) \  R^{R}(y, p_\perp) \ ; \ \ \ 
R^{R}(y, p_\perp) = \int_0^{\delta_{\rm max}(y)} \! \! \! \! \! \! \! \dd{\delta} \, \, \Phat_R(x) \, \frac{\dd\sigma_{\mathrm{pp}}^{h}(y+\delta, p_\perp)/\dd y  \dd p_\perp}{\dd\sigma_{\mathrm{pp}}^{h}(y, p_\perp)/\dd y  \dd p_\perp} \, .
\ee
Here, $x \equiv e^\delta -1$ and $\delta_{\rm max}(y)= \min\left(\ln{2},y_{\mathrm{max}}-y\right)$ where $y_{\mathrm{max}} = \ln(\!\sqrts/p_\perp)$.

The sole effect of parton energy loss is encoded in Eq.~\eq{RpA-y-xh}. In addition to energy loss, $p_\perp$-broadening can be simply implemented by shifting $\vec{p}_\perp$ by the nuclear broadening $\Delta \vec{p}_\perp$ when expressing the double differential \pA\ cross section in terms of that in \pp, as was done in Ref.~\cite{Arleo:2013zua}. When  $|\Delta \vec{p}_\perp| \ll K_\perp$, nuclear broadening cannot resolve the dijet, and is thus transferred to the dijet as a whole. Consequently, all dijet constituents undergo the same {\it rotation} of `angle' $\Delta \vec{p}_\perp/E_{i}$. The fragmenting gluon and the tagged hadron respectively acquire a transverse momentum shift $\xh \, \Delta \vec{p}_\perp$ and $z \, \xh \, \Delta \vec{p}_\perp$. Taking into account the effect of $p_\perp$-broadening modifies \eq{RpA-y-xh} to 
\vspace{-1mm}
\be
\label{RpA-y-xh-complete}
R^{h}_{\rm pA}(y, p_\perp) \simeq \sum_R \ P_R(\xh) \  R^{R}(y, p_\perp) \ ; \ \ \ 
R^{R}(y, p_\perp) = \int_0^{\delta_{\rm max}(y)} \! \! \! \! \! \! \! \dd{\delta} \, \, \Phat_R(x) \int \frac{\dd \varphi}{2\pi} \frac{\frac{\dd\sigma_{\mathrm{pp}}^{h}(y+\delta+ \delta', \, |\vec{p}_\perp - z\,\xh \, \Delta \vec{p}_\perp^{\, R}|)}{\dd y \, \dd p_\perp}}{\frac{\dd\sigma_{\mathrm{pp}}^{h}(y, p_\perp)}{\dd y \, \dd p_\perp}} \, ,
\vspace{-1mm}
\ee
where $\Delta \vec{p}_\perp^{\, R}$ (of azimutal angle $\varphi$) is the broadening of the dijet in color state $R$, and the rapidity shift $\delta'$ induced by the broadening reads $\delta' \equiv \ln{({p_\perp}/{|\vec{p}_\perp - z \, \xh \, \Delta \vec{p}_\perp^{\, R}|})}$.\footnote{The broadening $\Delta \vec{p}_\perp^{\, R}$ is given by, in average, $(\Delta \vec{p}_\perp^{{\,R}})^2 = \frac{N_c+C_R}{2N_c} (\hat{q} L - \hat{q}_{_{\rm p}} L_{_{\rm p}})$. The color factor arises from the average dijet color charge in the nucleus $(N_c+C_R)/2$ (since the dijet enters the nucleus with charge $N_c$ and exits with charge $C_R$) and after normalization by the gluon charge $N_c$ (recall that $\qhat$ is the {\it gluon} transport coefficient).}

\begin{figure}[t]
\centering
\includegraphics[width=4.8cm,height=4cm]{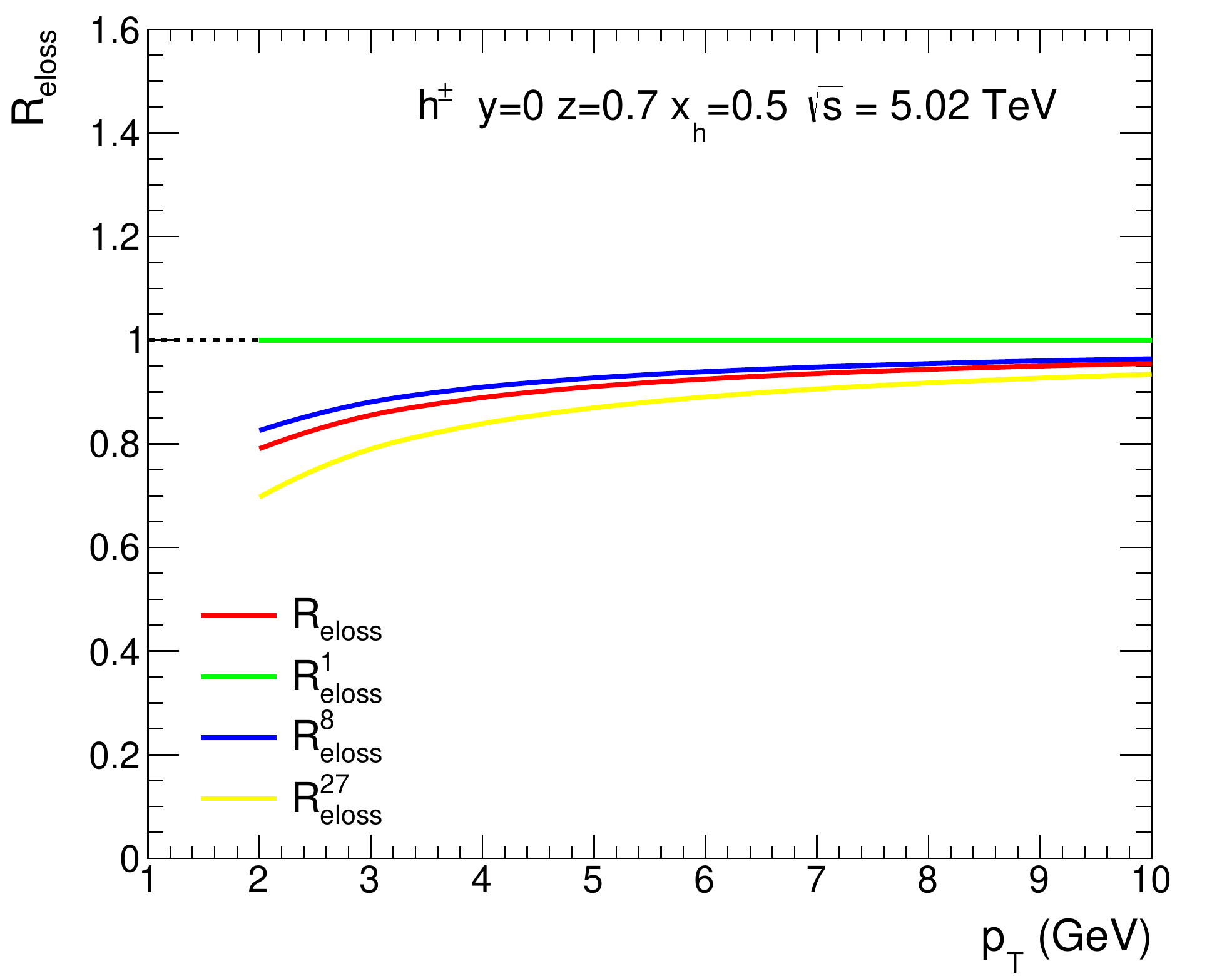}\includegraphics[width=4.8cm,height=4cm]{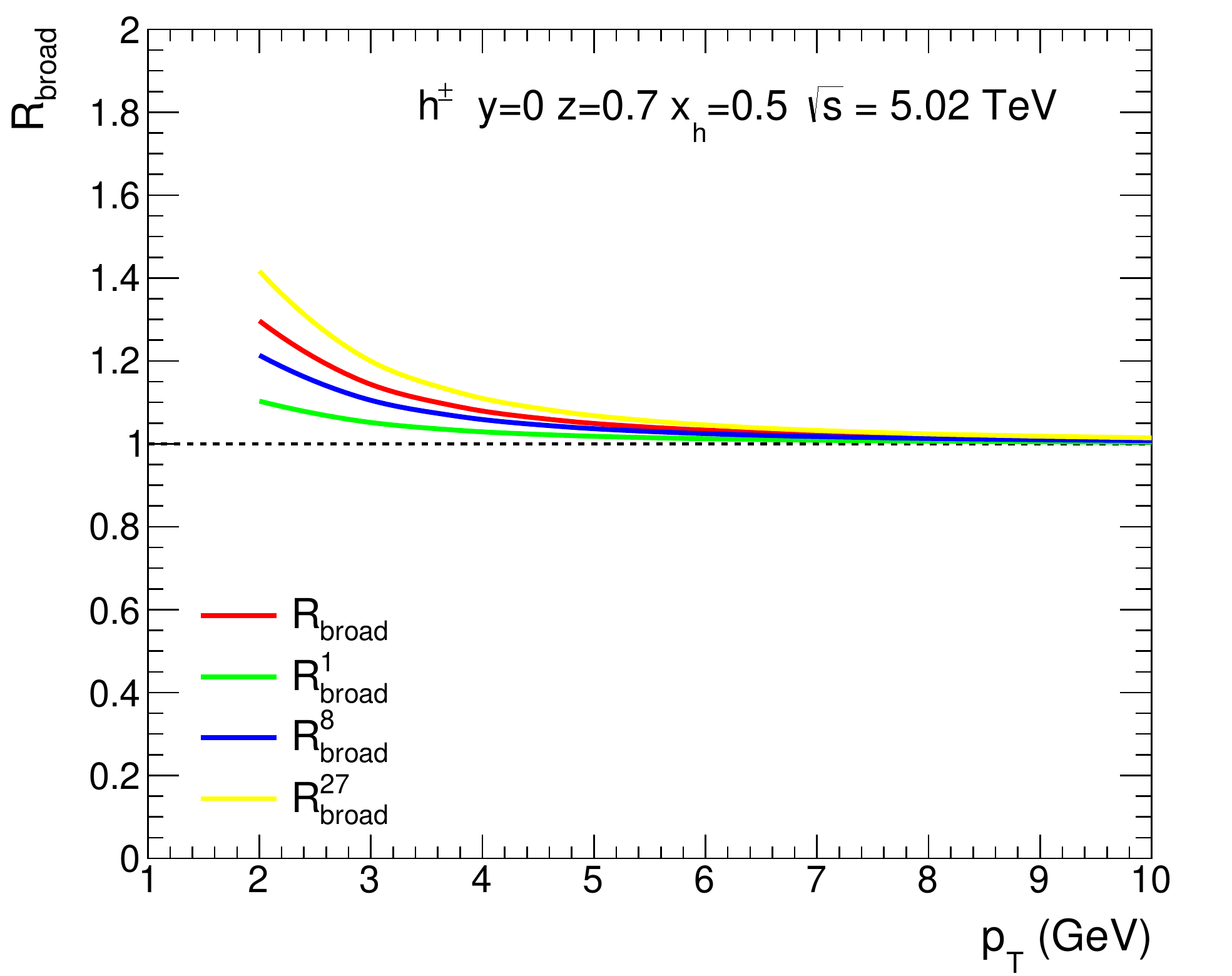}\includegraphics[width=4.8cm,height=4cm]{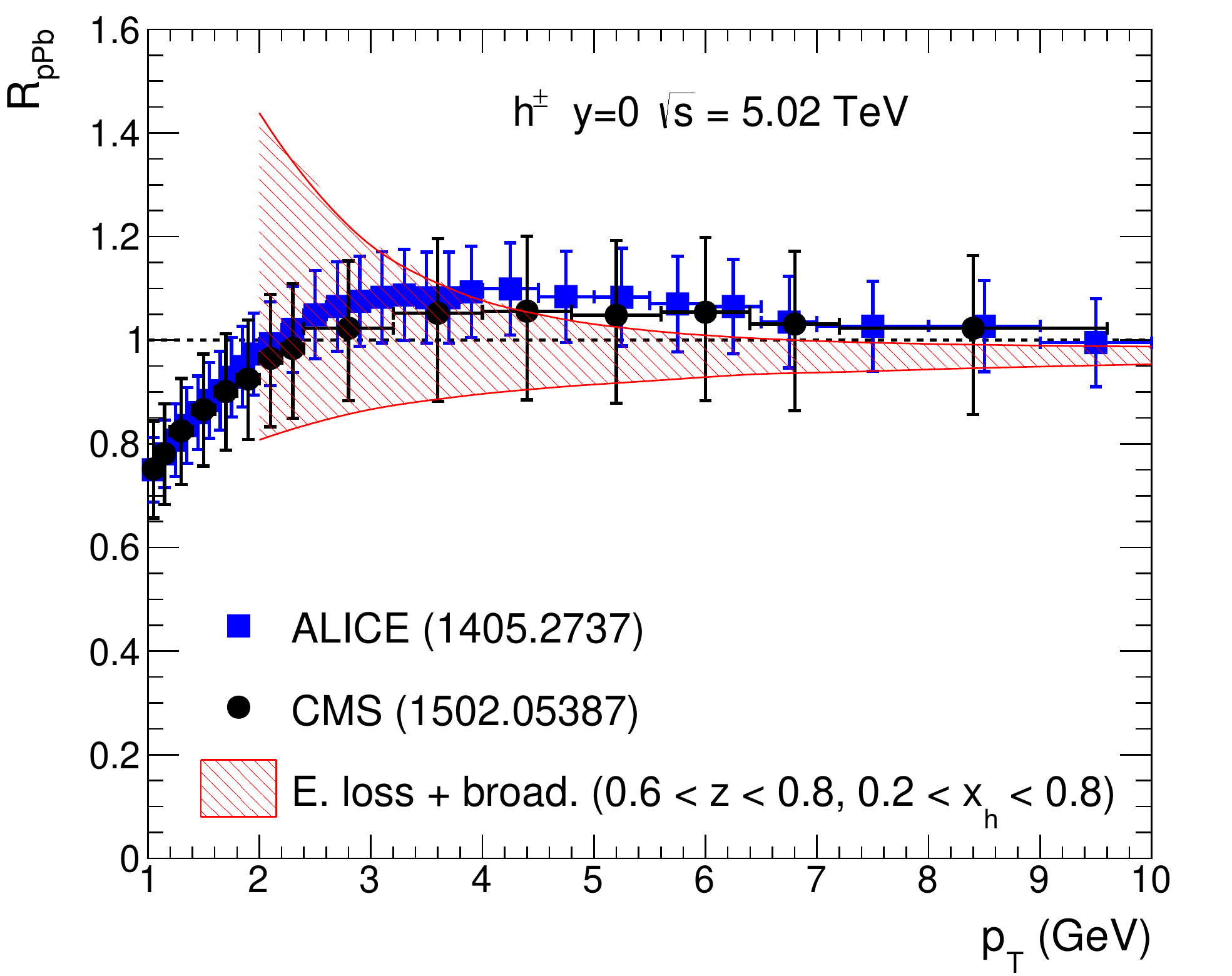}
\caption{Nuclear attenuation factor of charged hadrons in \pPb\ collisions at $\sqrts=5$~TeV including the effect of coherent energy loss only (left) and nuclear broadening only (middle) for different dijet color representations, and including both effects (right) as predicted in the model (band) and measured by ALICE~\cite{Abelev:2014dsa} and CMS~\cite{Khachatryan:2015xaa}.}
\label{RpA-eloss-broad}
\end{figure}

In order to compute $R^{h}_{\rm pPb}$ for charged hadron production at the LHC, we determine $\dd\sigma_{\mathrm{pp}}^{h}/\dd y \, \dd p_\perp$ from a fit to 7~TeV \pp\ data. The fragmentation variable $z$ and the dijet momentum fraction $\xh$ are chosen as $z=0.7$ and $\xh=0.5$ in average (based on estimates from perturbative calculations) but allowed to vary in the range $0.6 < z < 0.8$ and $0.2 < \xh < 0.8$.

We define the attenuation factors $R_{\mathrm{eloss}}$ and $R_{\mathrm{broad}}$ obtained from \eq{RpA-y-xh-complete} and corresponding respectively to the effect of energy loss only, and of nuclear broadening only. In Fig.~\ref{RpA-eloss-broad} (left) we compute $R^R_{\mathrm{eloss}}$ in \pPb\ collisions at $\sqrts=5$~TeV, for the accessible representations $R$ of the parent dijet, as well as for the color average $R_{\rm eloss} = \sum_R P_R(\xh)\, R^{R}_{\rm eloss}$, for $z=0.7$ and $\xh=0.5$. As can be seen, energy loss leads to a moderate suppression at low $\pt$ for $R={\bf 8}$ and slightly more pronounced for $R={\bf 27}$ due to the larger Casimir factor. (There is no coherent energy loss in the singlet channel $R={\bf 1}$.) An opposite trend is reported in Fig.~\ref{RpA-eloss-broad} (middle) regarding $R^R_{\mathrm{broad}}$: the higher the color representation $R$, the stronger the enhancement at small $\pt$. Energy loss and nuclear broadening effects thus tend to balance each other. In this respect, the {\it rapidity dependence} of $R^h_{\rm pPb}$ for charged hadrons (inclusive in $p_\perp \geq p_\perp^{_{\rm cut}} \sim 1$~GeV) would be interesting, since it is expected to be mostly driven by energy loss. 
The attenuation factor $R^{h}_{\rm pPb}$ including both coherent energy loss and nuclear broadening is shown in Fig.~\ref{RpA-eloss-broad} (right). Although the theoretical uncertainty, resulting from the variation of $\qzero$, $\xh$ and $z$ on the one hand and from the shape of the \pp\ cross section on the other hand, is rather large, the model prediction is compatible with ALICE~\cite{Abelev:2014dsa} and CMS~\cite{Khachatryan:2015xaa} data. 

More details on the present approach will follow in a forthcoming publication.

\section{Outlook}

Investigating further coherent energy loss in \pA\ collisions, in addition to its theoretical interest, should also help disentangling hot from cold nuclear effects in nucleus-nucleus (\hi) collisions. Indeed, the sole effect of coherent energy loss in cold nuclear matter, when extrapolated to \hi\ collisions, has been shown to yield a significant $\jpsi$ nuclear suppression~\cite{Arleo:2014oha}. This stresses the need to fully understand quarkonium/hadron nuclear suppression in \pA\ collisions before extracting the additional {\it hot} effects expected in \hi\ collisions.

\end{document}